\documentclass[preprintnumbers,showpacs, showkeys, twocolumn,secnumarabic,amssymb,amsmath,amsfonts,aps, prl]{revtex4-1}
\usepackage[dvips]{graphicx}
\usepackage[latin1]{inputenc}
\usepackage{amsthm,framed,fancybox}
\usepackage{amsfonts}
\usepackage{t1enc}
\usepackage[latin1]{inputenc}
\usepackage{amscd}
\usepackage{amstext}
\usepackage{amsbsy}
\usepackage{amsopn}
\usepackage{upref}
\usepackage{enumerate}
\usepackage{endnotes}
\usepackage{mathrsfs}
\usepackage{color}
\usepackage{cancel}
\usepackage{array}
\usepackage{epsfig}
\usepackage{fontenc}
\usepackage{textcomp}
\usepackage{mathrsfs}
\usepackage{hyperref}
\usepackage{booktabs}

\everymath{\displaystyle}

\newcommand{\affA}{Aix-Marseille University, Campus de Luminy, case 907,
CNRS Centre de Physique Théorique, UMR 7223, 13288 Marseille Cedex 09, France}

\begin{document}


\title{Long-range resonant interactions in biological systems}

\author{Jordane Preto}
\email{preto@cpt.univ-mrs.fr}
\author{Marco Pettini}
\email{pettini@cpt.univ-mrs.fr}
\affiliation{\affA}

\begin{abstract}
The issue of retarded long-range resonant interactions between two molecules with oscillating dipole moments is reinvestigated within the 
framework of classical electrodynamics. By taking advantage of a theorem in complex analysis, we present a simple method to work 
out the frequencies of the normal modes, which are then used to estimate the interaction potential. The main results thus
found are in perfect agreement with several outcomes obtained from quantum computations. Moreover, when applied to a biophysical context, 
our findings shed new light on Fröhlich's theory of selective long-range interactions between biomolecules. In particular, at variance 
with a long-standing belief, we show that sizable resonant long-range interactions may exist only if the interacting system is out of thermal 
equilibrium.

\end{abstract}

\date{\today}


\maketitle

\textit{Introduction}---Living organisms are well-known to support many biochemical processes which, besides being 
extremely specific, seem to follow a precise time schedule. In this context, the motion of the molecules
can hardly be described on the basis of thermal fluctuations only. 
Most studied concepts to this effect refer to electromagnetic \textit{short-range} interactions \cite{israelachvili}, which leaves
\textit{long-range} interactions poorly investigated thus far. This is essentially because in biological
systems free ions of cell water tend to screen any electrostatic potential at a distance that usually does not exceed a few angstroms. 
However, this screening proves generally inefficient for interactions involving oscillating electric fields.
To this respect, in 1972, Fröhlich emphasized \cite{frohlich72} that two molecular systems which would exhibit large oscillating 
dipole moments could interact via a \textit{long-range} potential, provided that the oscillation frequencies were roughly similar. 

Applied to biological systems, it was suggested that such resonant forces would have
a profound influence on the displacement of \textit{specific} biomolecular entities, and thus on the initiation of a particular 
cascade of chemical events. Later on, possible examples of such interactions were reported -- at the
cellular level -- between erythrocytes \cite{rowlands}. In parallel, it was shown that the membranes of these cells
have the ability to oscillate at frequencies of about $40$ GHz \cite{frohlich85}.
Moreover, since then \cite{cifra}, numerous evidences of electromagnetic long-range interactions between cells have been identified 
including yeast cells \cite{pohl}. 
Very recently, supported by experimental evidences of collective oscillations in biomacromolecules (see Refs. in \cite{article1}), possible experimental tests 
were proposed to assess whether such interactions could be relevant at the biomolecular level \cite{article1}.

Coming to Fröhlich's predictions, an intriguing result of his computations is the possibility of observing long-range resonant 
interactions even if the system of oscillating dipoles is at thermal equilibrium. This would occur when 
the retardation of the electric field mediating the interaction is relevant.
For large intermolecular distances $r$, the potential was given at thermal equilibrium as \cite{frohlich72,frohlich80}
\begin{equation}\label{frohlich_potential}
U(r) \propto \frac{1}{r^3} \left( \frac{1}{\varepsilon(\omega_+)} - \frac{1}{\varepsilon(\omega_-)} \right) + O\left(  \frac{1}{r^6} \right)
\end{equation}
where $\varepsilon(\omega_\pm)$ is the permittivity
of the medium at the normal frequencies $\omega_{\pm}$ of the interacting system.

Though Fröhlich's theory has been well received and developed for a long time by many authors \cite{pokornybook}, a thorough
derivation from first principles of his main results on resonant interactions, as for instance the potential given in Eq. \eqref{frohlich_potential}, is still lacking.
In anticipation of advanced experimental investigations \cite{article1},
the present paper is thus intended to be a more comprehensive and analytic account of this theory
and to put the mentioned interactions at their right place within the framework of electrodynamic forces.
To that purpose, we calculate the normal frequencies of a system of two interacting oscillating dipoles 
in an exact way, including field retardation. This is made possible by means of a theorem in 
complex analysis, the \textit{Lagrange inversion theorem}, that is detailed below. We then show
how these results can be used to estimate the interaction energy between two atoms on one hand, and two
oscillating dipoles on the other hand. In the former case, a connection with known QED results is provided and in
the latter one the connection with Fröhlich's theory is made. In particular, we prove that the form for the potential 
\eqref{frohlich_potential} is misleading, suggesting that \textit{long-range} interactions
exist only if the system is out of equilibrium.

\medskip

\textit{Theory}---Let us consider two molecules $A$ and $B$ with oscillating dipole moments $\boldsymbol{\mu}_A$ and $\boldsymbol{\mu}_B$. 
The equations of motion can be given in general terms as 
\begin{equation}\label{system0}
\left \{
\begin{array}{l}
 \ddot{\boldsymbol{\mu}}_A + \gamma_A \dot{\boldsymbol{\mu}}_A + \omega_A^2 \boldsymbol{\mu}_A = 
\zeta_A\boldsymbol{E}_B(\boldsymbol{r}_A,t) + \boldsymbol{f}_A(\boldsymbol{\mu}_A,t) \vspace{3mm} \\
 \ddot{\boldsymbol{\mu}}_B  + \gamma_B \dot{\boldsymbol{\mu}}_B +  \omega_B^2 \boldsymbol{\mu}_B = 
\zeta_B \boldsymbol{E}_A(\boldsymbol{r}_B,t) + \boldsymbol{f}_B(\boldsymbol{\mu}_B,t),
\end{array}
\right.
\end{equation}
where $\omega_{A,B}$ and $\gamma_{A,B}$ are the harmonic frequencies and damping coefficients of the dipoles. Here, the interaction takes 
place through the electric field $\boldsymbol{E}_{A, B}(\boldsymbol{r},t)$ generated by each molecule, located at $\boldsymbol{r}=\boldsymbol{r}_{B,A}$, 
while the coupling constants are given by $\zeta_{A} = Q_{A}^2/m_{A}$
where $Q_A$ and $m_A$ are the effective charge and mass involved in the dipole $A$, and similarly for $\zeta_B$.
Finally, $\boldsymbol{f}_A$ and $\boldsymbol{f}_B$ are functions accounting for possible anharmonic contributions, thermal noise
as well as possible external excitations.   

\medskip

\textit{Normal modes analysis}--- A possible way of estimating the interaction energy of the system (carried out in
the last part of this letter) is to use the normal frequencies of the associated harmonic conservative system given by
\begin{equation}\label{system}
\left \{
\begin{array}{l}
 \ddot{\boldsymbol{\mu}}_A + \omega_A^2 \boldsymbol{\mu}_A = \zeta_A\boldsymbol{E}_B(\boldsymbol{r}_A,t) \vspace{3mm} \\
 \ddot{\boldsymbol{\mu}}_B  + \omega_B^2 \boldsymbol{\mu}_B = \zeta_B \boldsymbol{E}_A(\boldsymbol{r}_B,t).\\
\end{array}
\right.
\end{equation}

The normal frequencies are defined as the frequencies $\omega$ such that
$
\boldsymbol{\mu}_{A,B}(t) = \boldsymbol{\mu}_{A,B} e^{i \omega t}
$ are solutions of \eqref{system}.
In this context, the electric field
generated by a harmonic dipole is well-known from classical electrodynamics \cite{jackson} and one has the following matrix relation
\begin{equation}\label{field}
 \boldsymbol{E}_B(\boldsymbol{r}_A, t) = \boldsymbol{\chi}(r,\omega)\boldsymbol{\mu}_{B} e^{i \omega t},
\end{equation}
with a similar expression for $\boldsymbol{E}_A(\boldsymbol{r}_B, t)$. Here, $\boldsymbol{\chi}$ represents
the generalized susceptibility matrix (retarded Green function) of the electric field 
-- note that the spatial dependence
of $\boldsymbol{\chi}$ in \eqref{field} reduces simply to the distance $r \equiv |\boldsymbol{r}_A - \boldsymbol{r}_B|$ because the intermediate medium 
is supposed to be homogeneous and isotropic. In addition, to model an aqueous biological environment filled with
free moving ions, in what follows we will consider it as a dispersive 
dielectric medium of complex permittivity  $\varepsilon(\omega)$.

At this stage, it should be remarked that the generalized susceptibility of the electric field is analytic everywhere on the complex plane except on
an uncountable subset where it has a discontinuity. In a homogeneous dielectric, this subset is given by
$\{ \omega \in \mathbb{C} \  |\ \text{Im}(\omega\sqrt{\varepsilon(\omega)})=0 \}$ while $\boldsymbol{\chi}$ reads as \cite{jackson}
\begin{subequations}\label{electricsus}
\begin{gather}
\begin{array}{l}
\chi_{11}(r,\omega) = \chi_{22}(r,\omega)  = - \frac{e^{\pm i \omega \sqrt{\varepsilon(\omega)} r /c}}{\varepsilon(\omega) r^3} \cdot
\vspace{0.3cm}\\ 
\hspace{2.5cm} 
\left(  1 \mp \frac{ i\omega
\sqrt{\varepsilon(\omega)} r}{c}  -    \frac{\omega^2
\varepsilon(\omega) r^2}{c^2}  \right), \vspace{0.3cm}
\end{array} \\
\chi_{33}(r,\omega)  =  \frac{2e^{\pm i \omega \sqrt{\varepsilon(\omega)} r /c}}{\varepsilon(\omega) r^3}
\left(  1 \mp \frac{ i\omega
\sqrt{\varepsilon(\omega)} r}{c}   \right),
\end{gather}
\end{subequations}
and $\chi_{ij}(r,\omega)  =  0$ when $i \neq j$; the $\pm$ sign is attributed to positive or negative values of
$\text{Im}(\omega\sqrt{\varepsilon(\omega)})$, respectively. Let us also specify that the diagonal form of $\boldsymbol{\chi}$ is due to
the choice to set the $z$ axis along $r$. Finally, it should be stressed that for real values of $\omega$, each $\chi_{ii}$ is a complex number whose
imaginary part essentially accounts for the dissipation due to the field propagation \cite{landau}. Bearing in mind that in
computing normal frequencies, one drops dissipation effects,
only the real parts denoted by $\chi'_{ii}$ will be thus considered in what follows. 

Substituting into Eq. \eqref{system} the expected harmonic forms for $\boldsymbol{\mu}_{A,B}$ and 
Eq. \eqref{field}, one obtains a system of linear algebraic equations. The 
existence of the solutions is then ensured by the vanishing of the determinant
$(\omega_A^2 - \omega^2)(\omega_B^2 - \omega^2) -  \zeta_A\zeta_B (\chi_{ii}'(r,\omega))^2  = 0$.
This indicates two possible values for $\omega>0$ for each $i$, namely, $\omega_{i,+}$ and $\omega_{i,-}$ verifying
\begin{equation}\label{secdegre}
\omega_{i,\pm}^{2} =  \frac{ \omega_A^2 + \omega_B^2}{2} \pm \sqrt{\left( \frac{\omega_A^2 - \omega_B^2}{2}\right)^2
+ \zeta_A\zeta_B \left(\chi_{ii}'(r,\omega_{i,\pm})\right)^2}.
\end{equation}

At this point, it must be emphasized that each normal frequency is strongly dependent
on the proximity of the frequencies of the dipoles, as detailed below.

\medskip

\textit{Resonant case}---When $\omega_A \simeq \omega_0 \equiv \omega_B$, equation \eqref{secdegre} is readily simplified and
at large separations, one can expand $\omega_{i,\pm}$ around $\omega_0$ as
\begin{equation}\label{resonantapp}
\begin{array}{ll}
\omega_{i,\pm} =  \omega_0  + \phi_{i,\pm}(r,\omega_{i,\pm}), \vspace{0.2cm} \\ \hspace{1cm} \text{with} \ \phi_{i,\pm} \simeq  \pm
\frac{\sqrt{\zeta_A\zeta_B}\chi_{ii}'}{2\omega_0} - \frac{\zeta_A\zeta_B(\chi_{ii}')^{2}}{8\omega_0^3}. 
\end{array}
\end{equation}

The implicitness of the equation can then be solved by making use of the Lagrange
inversion theorem \cite{whittaker} : \textit{let $\mathscr{C}$ be a contour in the complex plane surrounding a point $a$, and 
let $\phi$ a function analytic inside
and on $\mathscr{C}$. If $t\in \mathbb{C}$ is such that the inequality $|t \hspace{0.5mm} \phi(z)|< |z-a|$
is satisfied for all $z$ on the perimeter of $\mathscr{C}$, then the equation $\omega = a + t \phi(\omega)$ in $\omega$ has one
root inside $\mathscr{C}$ and any further function $g$ analytic inside and on $\mathscr{C}$ can be expanded
as a power series in $t$ by the formula}
\begin{equation}\label{lagrange}
g(\omega) =   g(a)  + \sum \limits_{n=1}^{\infty} \frac{t^n}{n!}
\frac{d^{n-1}}{da^{n-1}} \left[\frac{dg}{da} \left\{\phi(a) \right\}^n\right] .
\end{equation}

Identifying $t \phi(\omega)$ with $\phi_{i,\pm}(r,\omega)$ of equation \eqref{resonantapp}, one must
carefully choose the contour $\mathscr{C}$ out of the domain of discontinuity of $\chi_{ii}'$. 
On the other hand, since each $\chi_{ii}'$ (and so $\phi_{i}$) is a sum
of inverse power laws of $r$, we can assume that for large enough separation the inequality
$\phi_{i,\pm}(r,z) < |z -  \omega_0|$
holds for $z$ close enough to $\omega_0$ so that $\mathscr{C}$ may be still chosen in the domain of analyticity of $\chi_{ii}'$. Letting $g$ as 
the identity function and $n\le 2$, one finds after some algebra
\begin{equation}\label{resonantnormfreq}
\begin{array}{l}
\omega_{i,\pm}(r) \simeq  \omega_0   \pm  \frac{\sqrt{\zeta_A\zeta_B}\chi_{ii}'(r,\omega_0)}{2 \omega_0} + \vspace{0.3cm} \\ 
\hspace{2.9cm} 
\frac{\zeta_A\zeta_B}{2}
\frac{d}{d \omega} \left[\left(\frac{\chi_{ii}'(r,\omega)}{\omega + \omega_0} \right)^2 \right]_{\omega=\omega_0}, 
\end{array}
\end{equation}
to second order in $\chi_{ii}'$, $i = 1,2,3$.  The normal frequencies thus found are equal to the resonance frequency $\omega_0$
plus a shift due to the (effective) interaction. The first contribution of each frequency shift is proportional to the real part of the susceptibility matrix elements.
As detailed below, this term is responsible of the \textit{long-range} nature of the resonant interaction energy
as, according to \eqref{electricsus}, each $\boldsymbol{\chi}$-element reads as a polynomial in $1/r^\alpha$ with $\alpha \le 3$ (the dimension of physical space) : 
in the limit $r \ll c/\omega_{0}$ (near zone limit), it goes as $\pm 1/r^3$ with the intermolecular distance while it 
oscillates at longer distance (intermediate and far zone limits)
with a $\pm 1/r^2$ or even $ \pm 1/r$ envelope.

\medskip

\textit{Off-resonance case}---On the contrary, when $\omega_A\gg\omega_B$ (or similarly when $\omega_A \ll \omega_B$), 
one can use Eq. \eqref{secdegre} to approximate $\omega_{i,\pm}$ around $\omega_{A,B}$ as
$\omega_{i,\pm} \simeq  \omega_{A,B} \pm \zeta_A\zeta_B\left(\chi_{ii}'(r,\omega_{i,\pm})\right)^2 $ $/ \left[2 \omega_{A,B}(\omega_A^2 -
 \omega_B^2) \right]$
to second order in $\chi_{ii}'$. 
Again, for large $r$, the \textit{Lagrange inversion theorem} can be applied so that $t \phi$ of the theorem
corresponds to the last term of the equation. To compare the results with 
those of the resonant case, it is enough to apply Eq. \eqref{lagrange} up to $n=1$. Thus, the normal frequencies are
\begin{equation} \label{nonresonantnormfreq}
\omega_{i,\pm}(r) \simeq  \omega_{A,B} \pm
\frac{ \zeta_A \zeta_B \left(\chi_{ii}'(r,\omega_{A,B})\right)^2}{ 2 \omega_{A,B} \left(\omega_A^2 - \omega_B^2 \right)}.
\end{equation}
At variance with the resonant case, the frequency shifts associated with the unperturbed frequency $\omega_{A,B}$ are now proportional to 
$ \left(\chi_{ii}'\right)^2$ at first order. When $r\ll c/\omega_{A,B}$, this leads to a \textit{short-range} $\pm 1/r^6$ contribution
while at very long distances it oscillates decaying as  $\pm 1/r^2$.

\medskip

\textit{Interaction energy (atoms)}---Now that normal frequencies have been worked out, it is interesting to remark that the energy 
shifts $\textstyle{U_{\pm}(r) = \hbar \sum_{i} \left({\omega}_{i,\pm}(r) - \omega_{A,B}  \right)}$
are identical with the \textit{real} photon contributions that appears in the interaction energy of two two-level atoms 
in an excited state, as it can be found within the framework of QED. Since the atoms and the radiation field mediating the interaction can be considered as an 
ensemble of harmonic coupled oscillators, the correspondence is not so surprising if we add to this that normal modes and susceptibilities are the same for
classical and quantum oscillators. Thus, when $\omega_A \simeq \omega_0 = \omega_B$,
$U_{\pm}$ computed from Eq. (\ref{resonantnormfreq}) are equal
to the energy shifts due to the interaction between two atoms with a common transition frequency $\omega_0$ 
 \cite{stephen}.  Alternatively, 
in an off-resonance situation, one finds identical expressions for  $U_{\pm}$  computed from Eq. (\ref{nonresonantnormfreq}) and the 
energy shifts of two atoms with distinct transition frequencies, $\omega_A$ and $\omega_B$ respectively
(with $\omega_A \gg \omega_B$ here) \cite{mclone}. Readers interested 
in a more detailed investigation through the references mentioned above will identify 
$\zeta_A$ as $\omega_A |\tilde{\mu}_{A,i}|^2$ for all $\ i = 1, 2, 3$, where 
$\tilde{\mu}_{A,i}$ is the transition dipole moment of atom $A$ along
the $i^{\hspace{0.8mm} \text{th}}$ spatial coordinate (the same for atom $B$). The polarizability of each atom is then simply given by 
$$
\alpha_{A,B}(\omega) \equiv \frac{\zeta_{A,B}}{\omega_{A,B}^{\ 2} -\omega^2}.
$$

As a complement to these results, two important points should be mentioned.
\textit{First}, the above classical computation of normal frequencies does not allow to deduce the energy contribution due to 
\textit{virtual} photons that accounts for the interaction between the ground states of the atoms.
This is confirmed by the fact that this contribution which is always present whether 
the system is excited or not, was shown to arise purely from vacuum 
fluctuations \cite{power}. \textit{Secondly}, if we refer to recent quantum calculations \cite{sherkunov}, the \textit{real} photon contribution of the 
interaction energy in the case $\omega_A \neq \omega_B$ would be noticeably different from the one suggested in the present paper 
as well as in the references mentioned in \cite{mclone}. This result is distinguished by the presence of $|\chi_{ii}|^2$ instead of
$(\chi_{ii}')^2$ in Eq. \eqref{nonresonantnormfreq}, that leads to a spatially monotonic -- instead of oscillating -- potential.
Nevertheless, deeeper theoretical investigations \cite{haakh} revealed that the monotonic potential holds true on time scales much larger than the spontaneous
decay time of the excited atom(s). In the opposite case, the oscillating potential is obtained, thus recovering the above mentioned quantum-classical correspondence 
\cite{QED}.

\medskip

\textit{Interaction energy (real dipoles)}---Contrarily to the case of two atoms, the interaction between
real oscillating dipoles, \textit{i.e.}, molecules whose oscillating dipole moments are not due to electron motions but rather to conformational oscillations,
have not been given much attention in the literature. As mentioned in the Introduction, a remarkable exception is given by Fröhlich who tackled this
issue forty years ago in a biophysical context. 
At that time, he 
emphasized in particular that resonant long-range interactions may occur between two harmonic dipoles even if none of the normal modes
is excited beyond thermal equilibrium. However, we show that this statement is actually incorrect. 

To clarify, let us consider the system of dipoles $A$ and $B$ in thermal 
equilibrium and suppose as an example that 
$\hbar \omega_{A,B}\ll k_B T$ so that classical effects are dominant \cite{classical}. 
The interaction energy is then given by the difference of free energy of the coupled and uncoupled systems, \textit{i.e.},
$\textstyle{U(r) =  F(r) - F(\infty) = -k_B T \ln \left[Z(r)/Z(\infty) \right]}$,
where $Z(r)$ is the partition function of the system when the dipoles
are separated by a distance $r$. Using Boltzmann distribution
for each normal mode, one can easily show that the interaction energy has the following form \cite{frohlich80}
\begin{equation}\label{free_energy}
U(r) =  k_B T \sum \limits_i \ln \left[ \frac{{\omega}_{i,+}(r) {\omega}_{i,-}(r)}{\omega_A \omega_B} \right].
\end{equation}

Taking $\omega_A \simeq \omega_0 \equiv \omega_B$, the result exposed by Fröhlich is obtained by substituting 
$\omega_{i,\pm}(r)$ and  $\omega_{i,-}(r)$ for their \textit{implicit} form \eqref{resonantapp}. 
Then, at first order, $U$ reads as 
$$
U(r)  =  \frac{k_B T\sqrt{\zeta_A \zeta_B}}{2\omega_0^2} \sum \limits_i \left\{\chi_{ii}'(r,\omega_{i,+}) - \chi_{ii}'(r,\omega_{i,-}) \right\}.
$$

In addition, Fröhlich only considered the limit $r \ll  c / \omega_0$ together with the condition
$\varepsilon \simeq \varepsilon' \equiv \text{Re}(\varepsilon)$ 
(non absorbing medium in the considered frequency range). From Eq. \eqref{electricsus}, 
one has in this case :
\begin{equation}\label{frohlich_energy}
U(r)  \simeq  \frac{k_B T \sqrt{\zeta_A\zeta_B}}{2\omega_0^2}   \frac{1}{ r^3} \sum \limits_i
\sigma_i \left\{ \frac{1}{\varepsilon'(\omega_{i,+})} - \frac{1}{\varepsilon'(\omega_{i,-})} \right\},
\end{equation}
with $\sigma_1 = \sigma_2 = -1$ and $\sigma_3 = 2$. At this stage, Fröhlich claimed that the above $1/r^3$ form for $U$
may account for the existence of long-range resonant interactions in thermal equilibrium provided  
$\varepsilon'(\omega_{i,+}) \neq \varepsilon'(\omega_{i,-})$. 
However, as already mentioned, this form
arises simply since the implicitness has been not resolved yet.
Hence, by using the \textit{Lagrange inversion theorem} and by substituting, for example, $g$ of equation \eqref{lagrange}
for $1/\varepsilon'$, one get immediately that the terms in curly brackets in Eq. \eqref{frohlich_energy}
vanish at first order. As a result, expansion to second order shows that $U$ is actually proportional to $1/r^6$. 
To compute the complete form for $U$, one then needs to come back to Eq. \eqref{free_energy}. Using the \textit{explicit}
form of the normal frequencies derived above \eqref{resonantnormfreq}, one easily obtains
\begin{equation}
U(r) =  - \frac{3 k_B T \zeta_A\zeta_B}{\omega_0^{\hspace{0.7mm}4}[\varepsilon'(\omega_0)]^{\hspace{0.1mm} 2}}   \frac{1}{ r^6}
\left\{ 1 + \omega_0   \frac{d\ln[\varepsilon'(\omega_0)]}{d\omega_0} \right\},
\end{equation}

where the ability of the potential to be attractive or repulsive depends on the derivative of $\varepsilon'$ at $\omega_0$,
representative of the dispersive properties of the medium. 

In the end, it goes without saying that the long-range contributions that appear in the expression 
of the normal frequencies at resonance, Eq. \eqref{resonantnormfreq}, 
cancel each other since the energy of both modes is given by Boltzmann distribution; 
this happens independently of considering retardation effects.
By inference, long-range resonant 
interactions between two dipoles (biomolecules) may occur only beyond thermal equilibrium provided that the excitation of one normal 
mode is statistically ``favored'' compared to the other.
In this case, anharmonicity of 
the dipoles, as mentioned in commenting Eqs. \eqref{system0} would allow energy redistribution among normal modes whereas energy supply
could be essential to maintain a high degree of excitation despite energy losses. In a biological context this energy supply may
be attributed to environmental metabolic activity. 
This scenario was depicted by Fröhlich in general terms. He showed that a set of
coupled normal modes can undergo a condensation phenomenon characterized by the emerging of the mode of \textit{lowest} frequency
containing, in the average, nearly all the energy supply \cite{frohlich68}. Here such a process would
result in an effective \textit{attractive} potential whose amplitude is dependent on the ``stored'' energy \cite{pokornybook}
(of course, the above given remarks still apply in the quantum case when $\hbar \omega_0 \gg k_B T$).
To conclude, it is worth noting that retardations effects at large $r$ bring about interactions with a $1/r$ dependence [last terms in Eqs. \eqref{electricsus}], 
\textit{i.e.}, of much longer range with respect to the interactions proposed by Fröhlich. This last result could be of utmost relevance for a deeper understanding
of the highly organized molecular machinery in living matter, as emphasized in the Introduction.

\textit{Acknowledgments}---We warmly thank H. R. Haakh, P. J. Morrison, S. Spagnolo and M. Vittot for useful
comments and discussions.


\begin{thebibliography}{0}%
\makeatletter
\providecommand \@ifxundefined [1]{%
 \@ifx{#1\undefined}
}%
\providecommand \@ifnum [1]{%
 \ifnum #1\expandafter \@firstoftwo
 \else \expandafter \@secondoftwo
 \fi
}%
\providecommand \@ifx [1]{%
 \ifx #1\expandafter \@firstoftwo
 \else \expandafter \@secondoftwo
 \fi
}%
\providecommand \natexlab [1]{#1}%
\providecommand \enquote  [1]{``#1''}%
\providecommand \bibnamefont  [1]{#1}%
\providecommand \bibfnamefont [1]{#1}%
\providecommand \citenamefont [1]{#1}%
\providecommand \href@noop [0]{\@secondoftwo}%
\providecommand \href [0]{\begingroup \@sanitize@url \@href}%
\providecommand \@href[1]{\@@startlink{#1}\@@href}%
\providecommand \@@href[1]{\endgroup#1\@@endlink}%
\providecommand \@sanitize@url [0]{\catcode `\\12\catcode `\$12\catcode
  `\&12\catcode `\#12\catcode `\^12\catcode `\_12\catcode `\%12\relax}%
\providecommand \@@startlink[1]{}%
\providecommand \@@endlink[0]{}%
\providecommand \url  [0]{\begingroup\@sanitize@url \@url }%
\providecommand \@url [1]{\endgroup\@href {#1}{\urlprefix }}%
\providecommand \urlprefix  [0]{URL }%
\providecommand \Eprint [0]{\href }%
\@ifxundefined \urlstyle {%
  \providecommand \doi  [0]{\begingroup \@sanitize@url \@doi}%
  \providecommand \@doi [1]{\endgroup \@@startlink {\doibase
  #1}doi:\discretionary {}{}{}#1\@@endlink }%
}{%
  \providecommand \doi  [0]{doi:\discretionary{}{}{}\begingroup
  \urlstyle{rm}\Url }%
}%
\providecommand \doibase [0]{http://dx.doi.org/}%
\providecommand \Doi [0]{\begingroup \@sanitize@url \@Doi }%
\providecommand \@Doi  [1]{\endgroup\@@startlink{\doibase#1}\@@Doi}%
\providecommand \@@Doi [1]{#1\@@endlink}%
\providecommand \selectlanguage [0]{\@gobble}%
\providecommand \bibinfo  [0]{\@secondoftwo}%
\providecommand \bibfield  [0]{\@secondoftwo}%
\providecommand \translation [1]{[#1]}%
\providecommand \BibitemOpen [0]{}%
\providecommand \bibitemStop [0]{}%
\providecommand \bibitemNoStop [0]{.\EOS\space}%
\providecommand \EOS [0]{\spacefactor3000\relax}%
\providecommand \BibitemShut  [1]{\csname bibitem#1\endcsname}%
\end{thebibliography}%


\begin{thebibliography}{1}
\bibitem{israelachvili} J. N. Israelachvili, Quart. Rev. Biophys. \textbf{6}, 341 (1974).
\bibitem{frohlich72} H. Fröhlich, Phys. Lett. A\textbf{39}, 153 (1972).
\bibitem{rowlands} S. Rowlands, L. S. Sewchand, R. E. Lovlin, J. S. Beck and E. G. Enns, Phys. Lett. A{\bf 82}, 436 (1981); 
S. Rowlands, L. S. Sewchand, and E. G. Enns, Phys. Lett. A{\bf 87}, 256 (1982).
\bibitem{frohlich85} H. Fröhlich, Phys. Lett. A\textbf{110}, 480 (1985).
\bibitem{cifra} See the review paper :  M. Cifra, J. Z. Fields and A. Farhadi, Prog. Biophys. Mol. Biol. {\bf 105}, 223 (2011).
\bibitem{pohl} H. Pohl, J. Biol. Phys. \textbf{8}, 45 (1980).

\bibitem{article1} J. Preto, E. Floriani, I. Nardecchia, P. Ferrier and M. Pettini, Phys. Rev. E (2012), in press; arXiv/1201.2607.


\bibitem{frohlich80} H. Fröhlich, Adv. Electron. Electron Phys. \textbf{53}, 85 (1980).
\bibitem{pokornybook} J. Pokorn\'y and T. M. Wu, {\it Biophysical Aspects of Coherence and Biological Order}, (Springer, Berlin, 1998).

\bibitem{frohlich68} H. Fröhlich, Int. J. Quantum Chem. \textbf{2}, 641 (1968).
\bibitem{jackson} J. D. Jackson, {\it Classical Electrodynamics} (John Wiley \& Sons, Inc., Sixth Printing, 1967), p. 268-273.
\bibitem{landau} L. D. Landau and E. M. Lifshitz, {\it Statistical Physics}, (Pergamon Press, NY, 1980).
\bibitem{whittaker}  E. T. Whittaker and G. N. Watson, {\it A Course of Modern Analysis}, (Cambridge University Press, 
Fourth Edition 1927), p. 131-133.


\bibitem{stephen} This is the well-known condition for exchange degeneracy for which
$U_{\pm}$ are associated with the approximate eigenfunctions
$\textstyle{\vert\psi_{\pm}\rangle =\frac{1}{\sqrt{2}}\left[ \vert e_A,g_B\rangle \mp \vert g_A,e_B\rangle\right]}$
respectively, where $\vert e_A,g_B\rangle $ or $\vert g_A,e_B\rangle $ refer to the states where just one atom $A$ or $B$ is 
excited in the absence of interaction. See for example:  M. J. Stephen, J. Chem. Phys. \textbf{40}, 669 (1964); 
A.D. McLachlan, Molecular Phys. \textbf{8}, 409 (1964),
in the case of two identical atoms.

\bibitem{mclone} Here, $U_\pm$ are associated with the approximate eigenfunctions 
$\vert\psi_+\rangle =\vert e_A,g_B\rangle$ or $\vert\psi_-\rangle =\vert g_A,e_B\rangle$.  See for example: R. R. McLone and E. A. Power, Proc. Roy. Soc. 
A\textbf{286}, 573 (1965);
G.-I. Kweon and N. M. Lawandy, Phys. Rev. A\textbf{47}, 4513 (1993), erratum: Phys. Rev. A\textbf{49}, 2205 (1994).

\bibitem{power} E. A. Power and T. Thirunamachandran, Phys. Rev. A\textbf{48}, 4761 (1993).
\bibitem{sherkunov} Y. Sherkunov, Phys. Rev. A\textbf{72}, 052703 (2005).
\bibitem{haakh} H. R. Haakh, J. Schiefele, C. Henkel, Proceedings of QFExt (2011); arXiv/1111.3748.
\bibitem{QED} 
Despite numerous efforts, it seems that the mentioned monotonic potential 
is not derivable from classical arguments similar to those exposed here 
when dissipation is involved phenomenologically as it is 
the case when writing equations \eqref{system0}.
 
\bibitem{classical}  Note that at physiological temperature, \textit{i.e.}, $T \sim 35\text{°C}$, this condition is fulfilled for a wide range of frequencies as
$k_B T/ \hbar \simeq 2.10^{13} \ \text{s}^{-1}$.





\end{thebibliography}
\end{document}